\documentclass[aps,preprint,superscriptaddress]{revtex4-1}
\usepackage{amsmath,bm,amssymb,amsfonts}
\usepackage{dcolumn}
\usepackage{graphicx}
\usepackage{latexsym}
\usepackage{epsfig}
\usepackage{multirow}

\usepackage{color}

\newcommand{\vect}[1]{\mathbf{#1}}

\begin{document}

\title{Topological states in A15 superconductors}

\author{Minsung \surname{Kim}}
\affiliation{Ames Laboratory -- U.S. Department of Energy, Iowa State University, Ames, Iowa 50011, USA}
\author{Cai-Zhuang \surname{Wang}}
\affiliation{Ames Laboratory -- U.S. Department of Energy, Iowa State University, Ames, Iowa 50011, USA}
\author{Kai-Ming \surname{Ho}}
\affiliation{Ames Laboratory -- U.S. Department of Energy, Iowa State University, Ames, Iowa 50011, USA}
\affiliation{Department of Physics and Astronomy, Iowa State University, Ames, Iowa 50011, USA}

\date{\today}

\begin{abstract}
Superconductors with the A15 structure are prototypical type-II $s$-wave superconductors 
which have generated considerable interest in early superconducting material history. 
However, the topological nature of the electronic structure remains unnoticed so far.
Here, using first-principles band structure calculations based on density functional theory, 
we show that the A15 superconductors (Ta$_3$Sb, Ta$_3$Sn, and Ta$_3$Pb) 
have nontrivial band topology in the bulk electronic band structures, 
leading to the formation of topological surface states near the Fermi energy.
Due to the bulk superconductivity, 
the proximity effect in the topological surface states 
would induce topological superconductivity even without 
heterostructure of a topological insulator and an s-wave superconductor.
Our results indicate that the A15 superconductors are promising candidates for the realization of 
the topological superconductivity and the Majorana fermion.
\end{abstract}


\date{\today}

\maketitle


Topological phase of a matter has been 
a central theme of recent condensed matter physics~\cite{hasan2010,qi2011}. 
An intriguing aspect of the topological states is that 
they are promising platforms to realize novel excitations such as Majorana fermions 
in condensed matter systems
whereas they are elusive in high-energy physics~\cite{majorana1937,read2000,fu2008,volovik2009,sau2010,alicea2010,linder2010,wimmer2010,lutchyn2010,oreg2010}.
One of the propositions to realize Majorana fermions is making heterostructures 
between strong topological insulators and $s$-wave superconductors, 
where the proximity-induced superconductivity in the topological surface states 
is analogous to that in a spinless $p_x + i p_y$ superconductor 
that is expected to have Majorana bound states at the vortices~\cite{fu2008,read2000}.
A straightforward extension of the idea would be to search for conventional $s$-wave superconductors
which also have topological surface states~\cite{yan2013,xu2016,guan2016}.
The realization of topological superconductivity in a single material is advantageous
considering the possible complexity at the interfaces of heterostructures. 

A natural strategy for the search in this direction would be to examine 
existing superconductor material families in view of topological band theory.
In conventional Bardeen-Cooper-Schrieffer (BCS) superconductors,
the band structure in the normal phase is metallic 
and sizable density-of-states exists at the Fermi level. 
In contrast, the topological characterization of the band structure requires
a band gap in order to define a topological invariant of the occupied band manifold
since the band topology is defined using continuous deformation of the Hamiltonian without closing a gap. 
However, we note that it is still possible 
to define the band topology as long as there is a gap in the energy spectrum at each $k$-point (i.e., separate band manifolds).
While the band structure of the superconductor would have metallic bands at the Fermi level,
it is possible to have topological surface states in part of the Brillouin zone (BZ)
where the metallic bulk bands do not overlap with them for appropriate surface termination.
Thus, if we find a superconductor which has gaps at every $k$-point in the BZ 
with nontrivial band topology, it would be a potential candidate for the topological superconductor. 

Here, we show that superconductors with the A15 structure, 
which are representative metal-based superconductors mostly discovered 
from 1950's to 1970's~\cite{narlikar2014,buschow2005,hosono2017},
are promising candidates for topological superconductors.
Specifically, Ta$_3$Sb, Ta$_3$Sn, and Ta$_3$Pb are shown to 
have topological bulk band structures 
characterized by nontrivial $\mathbb{Z}_2$ invariants~\cite{fu2007titd}.
We find that the topological surface bands appear near the Fermi energy
as dictated by the bulk-boundary correspondence~\cite{hasan2010}.
Interestingly, since we have lower symmetry in the [001] surface, 
the topological surface states show non-helical spin texture unlike, e.g., Bi$_2$Se$_3$~\cite{hzhang2009,wzhang2010}.
We also discuss the electronic band structures of Nb compounds in the A15 family.


We performed first-principles electronic structure calculations 
based on density functional theory as implemented 
in Vienna \emph{ab initio} simulation package (VASP)~\cite{kresse1993,kresse1996}.
We employed the projector augmented-wave method~\cite{blochl1994}, 
and for the exchange correlation functional 
we adopted Perdew-Burke-Ernzerhof revised for solid (PBEsol)~\cite{perdew2008}.
We used a plane-wave basis set with the energy cutoff of 325 eV, and
$10\times10\times10$ $k$-point meshes were exploited.
The experimental lattice constants were used~\cite{luo1970,courtney1965,furuseth1964,guseva1982} 
except for the case where we could not find an experimental value.
For the surface state calculations, we used 21-layer-thick slab models in the [001] direction 
with sufficiently large vacuum regions ($\gtrapprox 23$ \AA) 
to prevent the spurious interactions between the periodic images.
The electronic structure was also checked and the symmetry property 
was analyzed using Quantum espresso package~\cite{giannozzi2009}.

\begin{figure}[]
\includegraphics[width=0.48\textwidth]{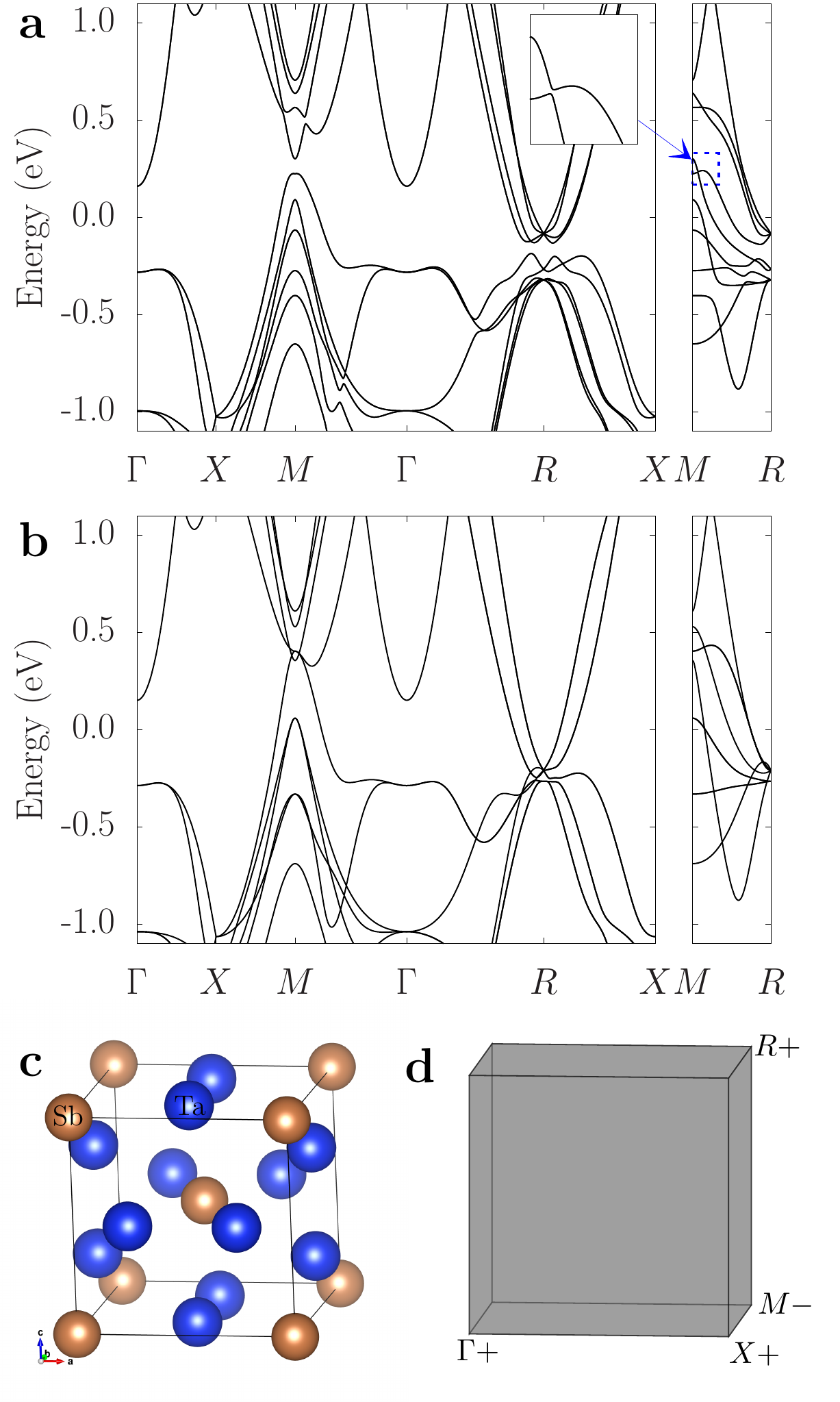}
\caption{\label{fig:atomic_band_str} Electronic band structure and atomic structure of Ta$_3$Sb.
The band structure along high-symmetry paths in the BZ (a) with and (b) without SOC.
(c) The atomic structure of Ta$_3$Sb.
(d) The parity products in the BZ.
The Fermi energy is set to 0.
}
\end{figure}

The superconductors with the A15 structure have cubic symmetry 
with the space group $Pm\overline{3}n$ \#223 (Fig.~\ref{fig:atomic_band_str}c).
They are intermetallic compounds with chemical formulae A$_3$B 
where A atoms lie at the cube face of the unit cell forming mutually orthogonal 
one-dimensional chains along edges, and B atoms constitute a bcc lattice.
The crystal structure belongs to a nonsymmorphic space group where for instance
we have a symmetry operation involving a screw axis, 
\begin{eqnarray}
\begin{pmatrix} 
x \\
y \\
z 
\end{pmatrix}
\rightarrow 
\begin{pmatrix} 
\frac{1}{2}-y \\
\frac{1}{2}+x \\
\frac{1}{2}+z 
\end{pmatrix}
=
\begin{pmatrix} 
0 & -1 & 0 \\
1 & 0 & 0 \\
0 & 0 & 1 	
\end{pmatrix}
\begin{pmatrix} 
x \\
y-\frac{1}{2} \\
z 	
\end{pmatrix}
+
\begin{pmatrix} 
0 \\
\frac{1}{2} \\
\frac{1}{2} 	
\end{pmatrix},
\end{eqnarray} 
with the lattice constant set to 1 for simplicity.
This operation is a 4-fold rotation around $z$ axis with respect to $(0,1/2,0)$ 
followed by a fractional translation along $z$ direction.
Accordingly, we have 4-fold rotational symmetry in the bulk electronic band structure 
although we do not have the strict 4-fold rotation in the crystal structure. 

The electronic band structure of the representative superconductor 
Ta$_3$Sb shows metallic feature with separate band manifolds, i.e., separation of the ``valence'' and the ``conduction'' bands 
in the whole BZ as shown in Fig.~\ref{fig:atomic_band_str}a.
We find that spin-orbit coupling (SOC) is essential to have the separation of the band manifolds (Fig.~\ref{fig:atomic_band_str}a, b).
The bands near the Fermi energy mostly have Ta $d$ character.
Since the crystal structure possesses the spatial inversion symmetry, 
the band topology of the valence bands can be calculated from the parities 
of the wavefunctions at the time-reversal invariant momenta (TRIM)~\cite{fu2007}.
The parity products are $+$,$+$,$-$, and $+$ for $\Gamma$, $3X$, $3M$, and $R$, respectively, 
which results in a strong topological phase 
with $(\nu_0;\nu_1 \nu_2 \nu_3) =(1;000)$ where $\nu_0$ is the strong $\mathbb{Z}_2$ index
and $\nu_1$, $\nu_2$, $\nu_3$ are the weak ones (Fig.~\ref{fig:atomic_band_str}d).

\begin{figure}[t]
\includegraphics[width=0.48\textwidth]{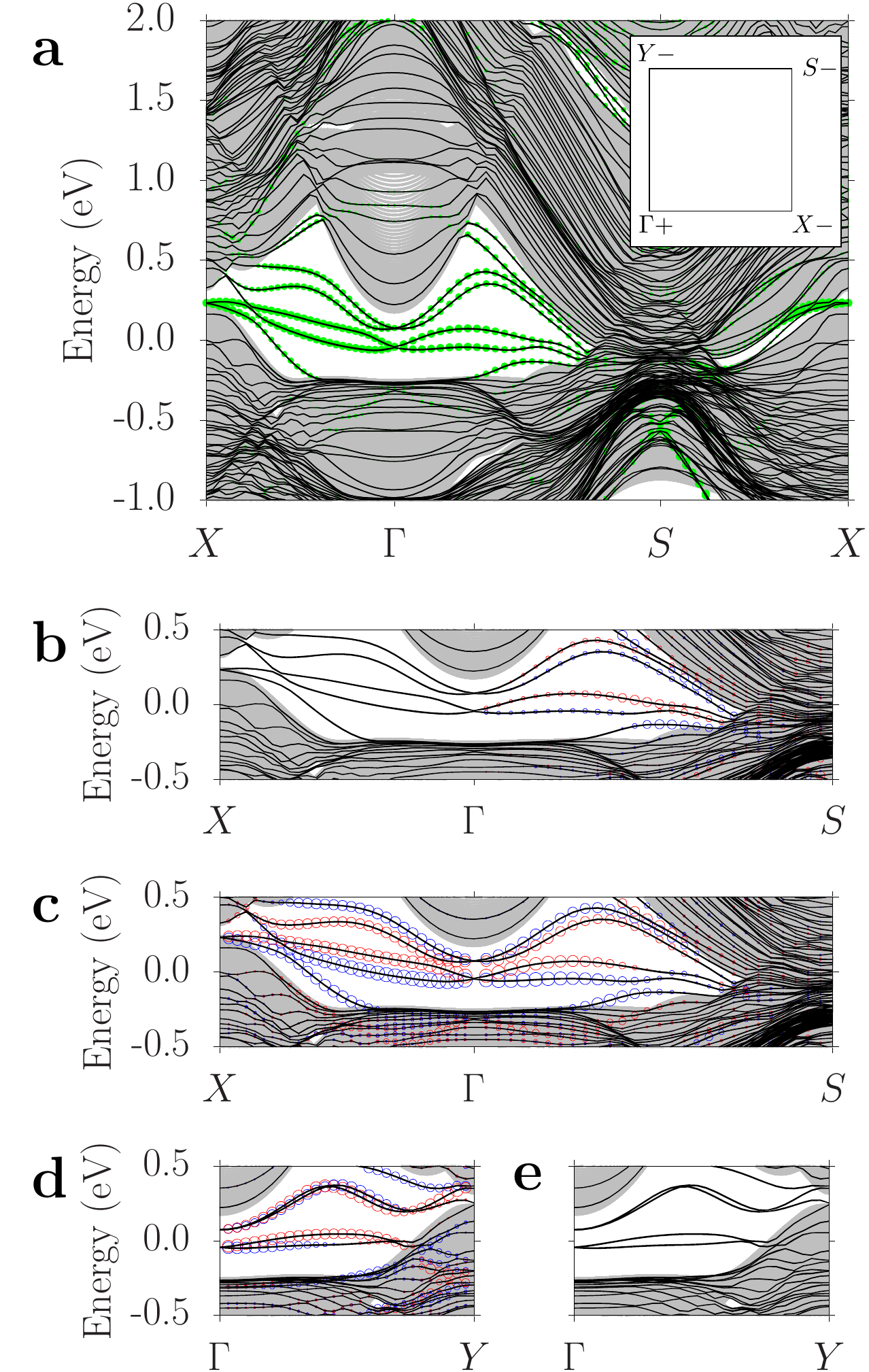}
\caption{\label{fig:surf_band} Surface states of Ta$_3$Sb in the [001] surface. 
(a) Electronic band structure along high-symmetry paths in the surface BZ.
Green circles denote the surface contribution. 
The inset shows the projected parity products at TRIM.
(b), (d) $x$-component of the spin angular momentum.
(c), (e) $y$-component of the spin angular momentum.
The red and blue circles denote the positive and negative values, respectively.
The grey regions correspond to the projected bulk states.
}
\end{figure}

One of the direct consequences of the nontrivial bulk band topology is 
the existence of the topological surface bands.
In the [001] surface of Ta$_3$Sb, we indeed find topological surface states as shown 
in Fig.~\ref{fig:surf_band}a.
The number (mod 2) of crossings between the surface states and a line connecting two TRIM 
in the gaps is dictated by the projected parity products.
For example, since we have different parities at $\Gamma$ and $X$ (i.e., $+$ at $\Gamma$ and $-$ at $X$), 
an odd number of crossings occurs along $\Gamma$--$X$ (Fig.~\ref{fig:surf_band}a), 
which characterizes topological surface states.
This confirms the bulk-boundary correspondence between the bulk topological invariants
and the surface state configurations.

\begin{figure}[]
\includegraphics[width=0.48\textwidth]{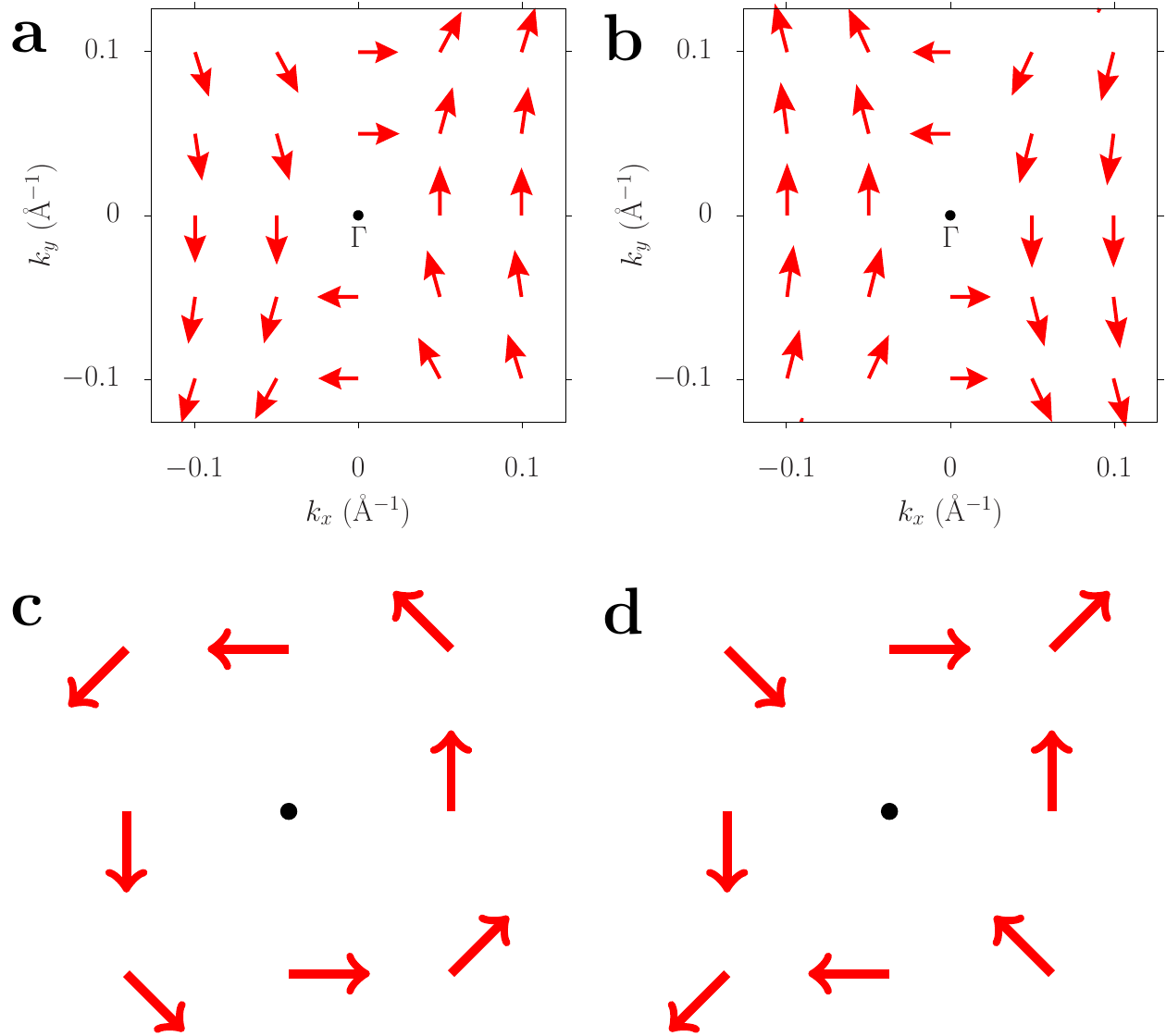}
\caption{\label{fig:am_explain} Spin angular momentum texture near $\Gamma$ in the [001] surface of Ta$_3$Sb.
The spin texture of the (a) lower and (b) upper bands.
The schematic illustration of (c) helical (Rashba-type) spin texture (d) linear Dresselhaus-type one. 
}
\end{figure}

The spin-polarized surface states have non-helical spin structure 
(as opposed to typical helical surface states, e.g., in Bi$_2$Se$_3$)
due to the reduced symmetry at the [001] surface (Fig~\ref{fig:surf_band}). 
In general, topological surface states are spin-polarized 
due to the SOC and the breaking of the inversion symmetry at the surface.
The specific spin configuration is determined by the relevant symmetry of the terminated surface.
Whereas we have the 4-fold symmetry in the bulk electronic bands,
we have only 2-fold symmetry in the [001] surface 
since the symmetry involving the fractional translation is broken.
The symmetry reduction at the surface is important 
for the qualitative understanding of the spin configuration.
To see this, we consider the effective Hamiltonian $H_{\mathrm{eff}} = \mathcal{A} k_y \sigma_x - \mathcal{B} k_x \sigma_y$ 
which is the general form under $C_{2v}$ (the relevant symmetry of the surface states near $\Gamma$) and the time-reversal symmetry 
up to linear order in $k$.
Here, $\sigma$'s are Pauli matrices that describe the spin degree of freedom.
$H_{\mathrm{eff}}$ gives an anisotropic Dirac cone
with the energy eigenvalues 
$E=\pm \lvert \vect{k} \rvert \sqrt{\mathcal{A}^2 \sin^2 \theta_{\vect{k}} + \mathcal{B}^2 \cos^2 \theta_{\vect{k}}}$,
in which $\cos \theta_{\vect{k}}=k_x / \lvert \vect{k} \rvert$ and $\sin \theta_{\vect{k}}=k_y / \lvert \vect{k} \rvert$.
When the local $k$ axis is rotated by $\pi/4$, the Hamiltonian takes the form  
$\alpha_R (k_y \sigma_x - k_x \sigma_y) + \alpha_D (k_x \sigma_x - k_y \sigma_y)$ 
with $\alpha_R = (\mathcal{A}+\mathcal{B})/2$ and $\alpha_D = (\mathcal{A}-\mathcal{B})/2$.
Here, the $\alpha_R$ term gives rise to 
helical (Rashba-type) spin texture (Fig.~\ref{fig:am_explain}c), 
and the $\alpha_D$ term leads to 
linear Dresselhaus-type one (Fig.~\ref{fig:am_explain}d)~\cite{mskim2016,stroppa2014}.
Indeed, our first-principles calculations show that 
the resulting spin configuration of the surface states near the Fermi energy around $\Gamma$ is 
a mixture of the two types of spin textures (Fig.~\ref{fig:am_explain}a). 
Note that under higher symmetry $C_{4v}$ or $C_{3v}$ 
we should have $\mathcal{A}=\mathcal{B}$, i.e., 
only the $\alpha_R$ (helical) term survives. 

Now we discuss the proximity induced superconductivity in the surface states.
Here, we consider the simplest case under the $C_{2v}$ symmetry and the time-reversal symmetry, 
where we have a single Dirac cone at the $\Gamma$ point.
If we substitute $k_x' = \frac{\mathcal{B}}{v} k_x$ and $k_y' = \frac{\mathcal{A}}{v} k_y$ with $v=\sqrt{\mathcal{A}^2+\mathcal{B}^2}$ in the effective Hamiltonian, 
the Hamiltonian becomes mathematically analogous to the isotropic helical Dirac cone described by $v (k_y' \sigma_x - k_x' \sigma_y)$.
The analysis can be given in a similar way with Ref.~\cite{fu2008} except for the spin helicity.
We consider the Cooper pairs tunneling into the surface states due to the proximity effect 
by introducing $H_{\mathrm{p}} = \Delta \psi_\uparrow^\dagger \psi_\downarrow^\dagger + \mathrm{H.c.}$ 
with $\Delta=\Delta_0 e^{i\phi}$.
Then the Bogoliubov--de Gennes (BdG) Hamiltonian is given by 
$H_{BdG} = \frac{1}{2} \Psi^\dagger \mathcal{H}_{BdG} \Psi$ where 
\begin{eqnarray}
\mathcal{H}_{BdG} = 
\begin{pmatrix} 
-iv (\sigma_x \partial_y - \sigma_y\partial_x) -\mu & \Delta \\
\Delta^\ast & iv (\sigma_x \partial_y - \sigma_y\partial_x) +\mu
\end{pmatrix},
\end{eqnarray} 
and $\Psi=(\psi_\uparrow,\psi_\downarrow,\psi_\downarrow^\dagger,-\psi_\uparrow^\dagger)^T$
with $\mu$ being the chemical potential.
By solving the BdG equation $\mathcal{H}_{BdG} \zeta = E \zeta$ for a vortex-like condition $\Delta=\Delta_0(r) e^{i\theta}$ in polar coordinates, the zero energy bound states can be found.
The solution for $\mu = 0$ case is simple and can be written as 
\begin{eqnarray}
\zeta (r,\theta) = 
\begin{pmatrix}
1 \\
0 \\
0 \\
-1
\end{pmatrix}	
\mathrm{exp}(-\int_{0}^{r} ds \frac{\Delta_0(s)}{v}),
\end{eqnarray}  
which represents the Majorana zero mode localized at the vortex.



\begin{figure}[]
\includegraphics[width=0.48\textwidth]{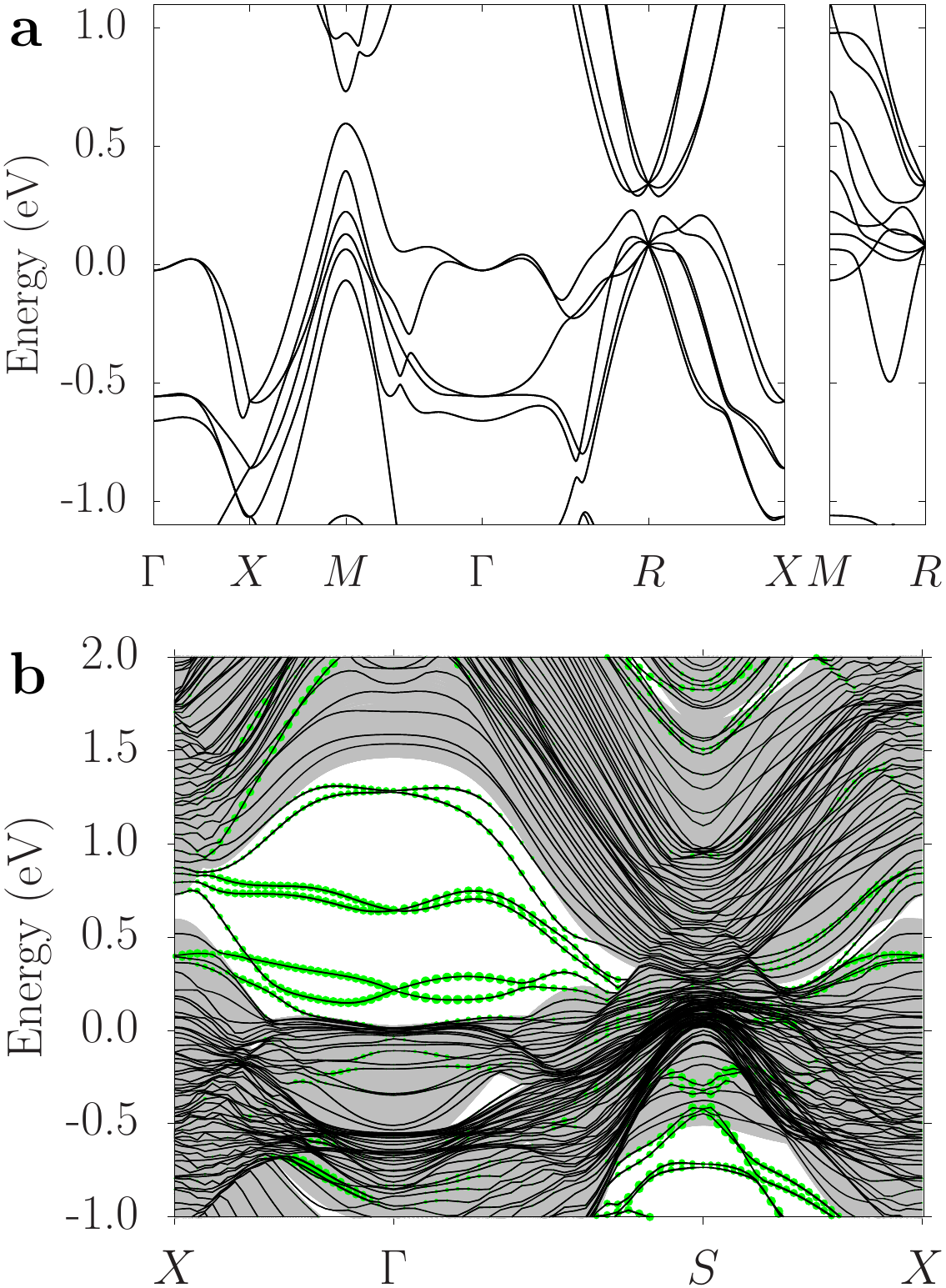}
\caption{\label{fig:ta3sn_band} Electronic structure of Ta$_3$Sn. 
(a) Bulk band structure along high-symmetry paths in Ta$_3$Sn. 
(b) Band structure of the [001] surface. Green circles denote the surface contribution.
The Fermi level is set to 0.
}
\end{figure}

The electronic band structure of Ta$_3$Sn shows qualitatively the same band topology.
Since we substitute Sb (Group VA in the periodic table) with Sn (Group IVA), 
the Fermi energy moves downwards below the surface states (Fig.~\ref{fig:ta3sn_band}).
The experimental superconducting transition temperatures ($T_c$) of Ta$_3$Sb
and Ta$_3$Sn are 0.7K and 8.3K, respectively~\cite{narlikar2014}.
We also find nontrivial band topology in Ta$_3$Pb (Pb also belongs to Group IVA) 
which has $T_c = 17$K~\cite{buschow2005}.
The compounds with Sn or Pb have the advantage that they have higher $T_c$
while electron doping would be needed to adjust the Fermi level in the surface.
The manipulation of the Fermi level in the surface layers could be possible using 
chemical substitution or liquid gating~\cite{ye2010,jeong2013}. 

Another variation of the composition would be the substitution of Ta with Nb 
since they belong to the same group in the periodic table.
However, we find that the valence and the conduction bands in Nb$_3$Sb and Nb$_3$Sn 
are not separated unlike the Ta compounds probably due to the weaker SOC strength of Nb. 
Thus, the $\mathbb{Z}_2$ bulk band topology is not properly defined in the Nb compounds.

In conclusion, we reported hitherto unnoticed topological phases in A15 superconductors, 
Ta$_3$Sb, Ta$_3$Sn, and Ta$_3$Pb.
The topological band structure was characterized by the nontrivial strong $\mathbb{Z}_2$ topological invariant $\nu_0 = 1$.
Corresponding topological surface states appear in the [001] surface, 
and they show non-helical spin texture due to the reduced symmetry at the surface. 
The proximity induced superconductivity at the surface would give rise to 
the Majorana zero mode.
The topological surface bands of the proposed compounds could be experimentally verified 
using angle-resolved photoemission spectroscopy (ARPES)~\cite{hasan2010,hsieh2008}.
Our study shows that the A15 superconductors are promising candidates for the realization of the topological superconductivity and Majorana fermions, and potentially useful for topological quantum computation.

~\\
\begin{acknowledgments}
We thank Suk Bum Chung for fruitful discussions.
This work was supported by the U.S. Department of Energy (DOE), Office of Science, Basic Energy Sciences, Materials Sciences and Engineering Division. The research was performed at Ames Laboratory, which is operated for the U.S. DOE by Iowa State University under Contract No. DE-AC02-07CH11358. Computations were performed through the support of the National Energy Research Scientific Computing Center, which is a DOE Office of Science User Facility operated under Contract No. DE-AC02-05CH11231.
\end{acknowledgments}


\begin{thebibliography}{36}%
\makeatletter
\providecommand \@ifxundefined [1]{%
 \@ifx{#1\undefined}
}%
\providecommand \@ifnum [1]{%
 \ifnum #1\expandafter \@firstoftwo
 \else \expandafter \@secondoftwo
 \fi
}%
\providecommand \@ifx [1]{%
 \ifx #1\expandafter \@firstoftwo
 \else \expandafter \@secondoftwo
 \fi
}%
\providecommand \natexlab [1]{#1}%
\providecommand \enquote  [1]{``#1''}%
\providecommand \bibnamefont  [1]{#1}%
\providecommand \bibfnamefont [1]{#1}%
\providecommand \citenamefont [1]{#1}%
\providecommand \href@noop [0]{\@secondoftwo}%
\providecommand \href [0]{\begingroup \@sanitize@url \@href}%
\providecommand \@href[1]{\@@startlink{#1}\@@href}%
\providecommand \@@href[1]{\endgroup#1\@@endlink}%
\providecommand \@sanitize@url [0]{\catcode `\\12\catcode `\$12\catcode
  `\&12\catcode `\#12\catcode `\^12\catcode `\_12\catcode `\%12\relax}%
\providecommand \@@startlink[1]{}%
\providecommand \@@endlink[0]{}%
\providecommand \url  [0]{\begingroup\@sanitize@url \@url }%
\providecommand \@url [1]{\endgroup\@href {#1}{\urlprefix }}%
\providecommand \urlprefix  [0]{URL }%
\providecommand \Eprint [0]{\href }%
\providecommand \doibase [0]{http://dx.doi.org/}%
\providecommand \selectlanguage [0]{\@gobble}%
\providecommand \bibinfo  [0]{\@secondoftwo}%
\providecommand \bibfield  [0]{\@secondoftwo}%
\providecommand \translation [1]{[#1]}%
\providecommand \BibitemOpen [0]{}%
\providecommand \bibitemStop [0]{}%
\providecommand \bibitemNoStop [0]{.\EOS\space}%
\providecommand \EOS [0]{\spacefactor3000\relax}%
\providecommand \BibitemShut  [1]{\csname bibitem#1\endcsname}%
\let\auto@bib@innerbib\@empty
\bibitem [{\citenamefont {Hasan}\ and\ \citenamefont {Kane}(2010)}]{hasan2010}%
  \BibitemOpen
  \bibfield  {author} {\bibinfo {author} {\bibfnamefont {M.~Z.}\ \bibnamefont
  {Hasan}}\ and\ \bibinfo {author} {\bibfnamefont {C.~L.}\ \bibnamefont
  {Kane}},\ }\href {\doibase 10.1103/RevModPhys.82.3045} {\bibfield  {journal}
  {\bibinfo  {journal} {Rev. Mod. Phys.}\ }\textbf {\bibinfo {volume} {82}},\
  \bibinfo {pages} {3045} (\bibinfo {year} {2010})}\BibitemShut {NoStop}%
\bibitem [{\citenamefont {Qi}\ and\ \citenamefont {Zhang}(2011)}]{qi2011}%
  \BibitemOpen
  \bibfield  {author} {\bibinfo {author} {\bibfnamefont {X.-L.}\ \bibnamefont
  {Qi}}\ and\ \bibinfo {author} {\bibfnamefont {S.-C.}\ \bibnamefont {Zhang}},\
  }\href {\doibase 10.1103/RevModPhys.83.1057} {\bibfield  {journal} {\bibinfo
  {journal} {Rev. Mod. Phys.}\ }\textbf {\bibinfo {volume} {83}},\ \bibinfo
  {pages} {1057} (\bibinfo {year} {2011})}\BibitemShut {NoStop}%
\bibitem [{\citenamefont {Majorana}(1937)}]{majorana1937}%
  \BibitemOpen
  \bibfield  {author} {\bibinfo {author} {\bibfnamefont {E.}~\bibnamefont
  {Majorana}},\ }\href {\doibase 10.1007/BF02961314} {\bibfield  {journal}
  {\bibinfo  {journal} {Nuovo Cimento}\ }\textbf {\bibinfo {volume} {14}},\
  \bibinfo {pages} {171} (\bibinfo {year} {1937})}\BibitemShut {NoStop}%
\bibitem [{\citenamefont {Read}\ and\ \citenamefont {Green}(2000)}]{read2000}%
  \BibitemOpen
  \bibfield  {author} {\bibinfo {author} {\bibfnamefont {N.}~\bibnamefont
  {Read}}\ and\ \bibinfo {author} {\bibfnamefont {D.}~\bibnamefont {Green}},\
  }\href {\doibase 10.1103/PhysRevB.61.10267} {\bibfield  {journal} {\bibinfo
  {journal} {Phys. Rev. B}\ }\textbf {\bibinfo {volume} {61}},\ \bibinfo
  {pages} {10267} (\bibinfo {year} {2000})}\BibitemShut {NoStop}%
\bibitem [{\citenamefont {Fu}\ and\ \citenamefont {Kane}(2008)}]{fu2008}%
  \BibitemOpen
  \bibfield  {author} {\bibinfo {author} {\bibfnamefont {L.}~\bibnamefont
  {Fu}}\ and\ \bibinfo {author} {\bibfnamefont {C.~L.}\ \bibnamefont {Kane}},\
  }\href {\doibase 10.1103/PhysRevLett.100.096407} {\bibfield  {journal}
  {\bibinfo  {journal} {Phys. Rev. Lett.}\ }\textbf {\bibinfo {volume} {100}},\
  \bibinfo {pages} {096407} (\bibinfo {year} {2008})}\BibitemShut {NoStop}%
\bibitem [{\citenamefont {Volovik}(2009)}]{volovik2009}%
  \BibitemOpen
  \bibfield  {author} {\bibinfo {author} {\bibfnamefont {G.~E.}\ \bibnamefont
  {Volovik}},\ }\href {\doibase 10.1134/S0021364009170172} {\bibfield
  {journal} {\bibinfo  {journal} {JETP Letters}\ }\textbf {\bibinfo {volume}
  {90}},\ \bibinfo {pages} {398} (\bibinfo {year} {2009})}\BibitemShut
  {NoStop}%
\bibitem [{\citenamefont {Sau}\ \emph {et~al.}(2010)\citenamefont {Sau},
  \citenamefont {Lutchyn}, \citenamefont {Tewari},\ and\ \citenamefont
  {Das~Sarma}}]{sau2010}%
  \BibitemOpen
  \bibfield  {author} {\bibinfo {author} {\bibfnamefont {J.~D.}\ \bibnamefont
  {Sau}}, \bibinfo {author} {\bibfnamefont {R.~M.}\ \bibnamefont {Lutchyn}},
  \bibinfo {author} {\bibfnamefont {S.}~\bibnamefont {Tewari}}, \ and\ \bibinfo
  {author} {\bibfnamefont {S.}~\bibnamefont {Das~Sarma}},\ }\href {\doibase
  10.1103/PhysRevLett.104.040502} {\bibfield  {journal} {\bibinfo  {journal}
  {Phys. Rev. Lett.}\ }\textbf {\bibinfo {volume} {104}},\ \bibinfo {pages}
  {040502} (\bibinfo {year} {2010})}\BibitemShut {NoStop}%
\bibitem [{\citenamefont {Alicea}(2010)}]{alicea2010}%
  \BibitemOpen
  \bibfield  {author} {\bibinfo {author} {\bibfnamefont {J.}~\bibnamefont
  {Alicea}},\ }\href {\doibase 10.1103/PhysRevB.81.125318} {\bibfield
  {journal} {\bibinfo  {journal} {Phys. Rev. B}\ }\textbf {\bibinfo {volume}
  {81}},\ \bibinfo {pages} {125318} (\bibinfo {year} {2010})}\BibitemShut
  {NoStop}%
\bibitem [{\citenamefont {Linder}\ \emph {et~al.}(2010)\citenamefont {Linder},
  \citenamefont {Tanaka}, \citenamefont {Yokoyama}, \citenamefont {Sudb\o{}},\
  and\ \citenamefont {Nagaosa}}]{linder2010}%
  \BibitemOpen
  \bibfield  {author} {\bibinfo {author} {\bibfnamefont {J.}~\bibnamefont
  {Linder}}, \bibinfo {author} {\bibfnamefont {Y.}~\bibnamefont {Tanaka}},
  \bibinfo {author} {\bibfnamefont {T.}~\bibnamefont {Yokoyama}}, \bibinfo
  {author} {\bibfnamefont {A.}~\bibnamefont {Sudb\o{}}}, \ and\ \bibinfo
  {author} {\bibfnamefont {N.}~\bibnamefont {Nagaosa}},\ }\href {\doibase
  10.1103/PhysRevLett.104.067001} {\bibfield  {journal} {\bibinfo  {journal}
  {Phys. Rev. Lett.}\ }\textbf {\bibinfo {volume} {104}},\ \bibinfo {pages}
  {067001} (\bibinfo {year} {2010})}\BibitemShut {NoStop}%
\bibitem [{\citenamefont {Wimmer}\ \emph {et~al.}(2010)\citenamefont {Wimmer},
  \citenamefont {Akhmerov}, \citenamefont {Medvedyeva}, \citenamefont
  {Tworzyd\l{}o},\ and\ \citenamefont {Beenakker}}]{wimmer2010}%
  \BibitemOpen
  \bibfield  {author} {\bibinfo {author} {\bibfnamefont {M.}~\bibnamefont
  {Wimmer}}, \bibinfo {author} {\bibfnamefont {A.~R.}\ \bibnamefont
  {Akhmerov}}, \bibinfo {author} {\bibfnamefont {M.~V.}\ \bibnamefont
  {Medvedyeva}}, \bibinfo {author} {\bibfnamefont {J.}~\bibnamefont
  {Tworzyd\l{}o}}, \ and\ \bibinfo {author} {\bibfnamefont {C.~W.~J.}\
  \bibnamefont {Beenakker}},\ }\href {\doibase 10.1103/PhysRevLett.105.046803}
  {\bibfield  {journal} {\bibinfo  {journal} {Phys. Rev. Lett.}\ }\textbf
  {\bibinfo {volume} {105}},\ \bibinfo {pages} {046803} (\bibinfo {year}
  {2010})}\BibitemShut {NoStop}%
\bibitem [{\citenamefont {Lutchyn}\ \emph {et~al.}(2010)\citenamefont
  {Lutchyn}, \citenamefont {Sau},\ and\ \citenamefont
  {Das~Sarma}}]{lutchyn2010}%
  \BibitemOpen
  \bibfield  {author} {\bibinfo {author} {\bibfnamefont {R.~M.}\ \bibnamefont
  {Lutchyn}}, \bibinfo {author} {\bibfnamefont {J.~D.}\ \bibnamefont {Sau}}, \
  and\ \bibinfo {author} {\bibfnamefont {S.}~\bibnamefont {Das~Sarma}},\ }\href
  {\doibase 10.1103/PhysRevLett.105.077001} {\bibfield  {journal} {\bibinfo
  {journal} {Phys. Rev. Lett.}\ }\textbf {\bibinfo {volume} {105}},\ \bibinfo
  {pages} {077001} (\bibinfo {year} {2010})}\BibitemShut {NoStop}%
\bibitem [{\citenamefont {Oreg}\ \emph {et~al.}(2010)\citenamefont {Oreg},
  \citenamefont {Refael},\ and\ \citenamefont {von Oppen}}]{oreg2010}%
  \BibitemOpen
  \bibfield  {author} {\bibinfo {author} {\bibfnamefont {Y.}~\bibnamefont
  {Oreg}}, \bibinfo {author} {\bibfnamefont {G.}~\bibnamefont {Refael}}, \ and\
  \bibinfo {author} {\bibfnamefont {F.}~\bibnamefont {von Oppen}},\ }\href
  {\doibase 10.1103/PhysRevLett.105.177002} {\bibfield  {journal} {\bibinfo
  {journal} {Phys. Rev. Lett.}\ }\textbf {\bibinfo {volume} {105}},\ \bibinfo
  {pages} {177002} (\bibinfo {year} {2010})}\BibitemShut {NoStop}%
\bibitem [{\citenamefont {Yan}\ \emph {et~al.}(2013)\citenamefont {Yan},
  \citenamefont {Jansen},\ and\ \citenamefont {Felser}}]{yan2013}%
  \BibitemOpen
  \bibfield  {author} {\bibinfo {author} {\bibfnamefont {B.}~\bibnamefont
  {Yan}}, \bibinfo {author} {\bibfnamefont {M.}~\bibnamefont {Jansen}}, \ and\
  \bibinfo {author} {\bibfnamefont {C.}~\bibnamefont {Felser}},\ }\href@noop {}
  {\bibfield  {journal} {\bibinfo  {journal} {Nature Physics}\ }\textbf
  {\bibinfo {volume} {9}},\ \bibinfo {pages} {709} (\bibinfo {year}
  {2013})}\BibitemShut {NoStop}%
\bibitem [{\citenamefont {Xu}\ \emph {et~al.}(2016)\citenamefont {Xu},
  \citenamefont {Lian}, \citenamefont {Tang}, \citenamefont {Qi},\ and\
  \citenamefont {Zhang}}]{xu2016}%
  \BibitemOpen
  \bibfield  {author} {\bibinfo {author} {\bibfnamefont {G.}~\bibnamefont
  {Xu}}, \bibinfo {author} {\bibfnamefont {B.}~\bibnamefont {Lian}}, \bibinfo
  {author} {\bibfnamefont {P.}~\bibnamefont {Tang}}, \bibinfo {author}
  {\bibfnamefont {X.-L.}\ \bibnamefont {Qi}}, \ and\ \bibinfo {author}
  {\bibfnamefont {S.-C.}\ \bibnamefont {Zhang}},\ }\href {\doibase
  10.1103/PhysRevLett.117.047001} {\bibfield  {journal} {\bibinfo  {journal}
  {Phys. Rev. Lett.}\ }\textbf {\bibinfo {volume} {117}},\ \bibinfo {pages}
  {047001} (\bibinfo {year} {2016})}\BibitemShut {NoStop}%
\bibitem [{\citenamefont {Guan}\ \emph {et~al.}(2016)\citenamefont {Guan},
  \citenamefont {Chen}, \citenamefont {Chu}, \citenamefont {Sankar},
  \citenamefont {Chou}, \citenamefont {Jeng}, \citenamefont {Chang},\ and\
  \citenamefont {Chuang}}]{guan2016}%
  \BibitemOpen
  \bibfield  {author} {\bibinfo {author} {\bibfnamefont {S.-Y.}\ \bibnamefont
  {Guan}}, \bibinfo {author} {\bibfnamefont {P.-J.}\ \bibnamefont {Chen}},
  \bibinfo {author} {\bibfnamefont {M.-W.}\ \bibnamefont {Chu}}, \bibinfo
  {author} {\bibfnamefont {R.}~\bibnamefont {Sankar}}, \bibinfo {author}
  {\bibfnamefont {F.}~\bibnamefont {Chou}}, \bibinfo {author} {\bibfnamefont
  {H.-T.}\ \bibnamefont {Jeng}}, \bibinfo {author} {\bibfnamefont {C.-S.}\
  \bibnamefont {Chang}}, \ and\ \bibinfo {author} {\bibfnamefont {T.-M.}\
  \bibnamefont {Chuang}},\ }\href {\doibase 10.1126/sciadv.1600894} {\bibfield
  {journal} {\bibinfo  {journal} {Science Advances}\ }\textbf {\bibinfo
  {volume} {2}} (\bibinfo {year} {2016}),\ 10.1126/sciadv.1600894}\BibitemShut
  {NoStop}%
\bibitem [{\citenamefont {Narlikar}(2014)}]{narlikar2014}%
  \BibitemOpen
  \bibfield  {author} {\bibinfo {author} {\bibfnamefont {A.}~\bibnamefont
  {Narlikar}},\ }\href {https://books.google.com/books?id=3iYUDAAAQBAJ} {\emph
  {\bibinfo {title} {Superconductors}}}\ (\bibinfo  {publisher} {OUP Oxford},\
  \bibinfo {year} {2014})\BibitemShut {NoStop}%
\bibitem [{\citenamefont {Buschow}(2005)}]{buschow2005}%
  \BibitemOpen
  \bibfield  {author} {\bibinfo {author} {\bibfnamefont {K.}~\bibnamefont
  {Buschow}},\ }\href {https://books.google.com/books?id=N9mvytGEBtwC} {\emph
  {\bibinfo {title} {Concise Encyclopedia of Magnetic and Superconducting
  Materials}}},\ Advances in Materials Sciences and Engineering\ (\bibinfo
  {publisher} {Elsevier Science},\ \bibinfo {year} {2005})\BibitemShut
  {NoStop}%
\bibitem [{\citenamefont {Hosono}\ \emph {et~al.}(2017)\citenamefont {Hosono},
  \citenamefont {Yamamoto}, \citenamefont {Hiramatsu},\ and\ \citenamefont
  {Ma}}]{hosono2017}%
  \BibitemOpen
  \bibfield  {author} {\bibinfo {author} {\bibfnamefont {H.}~\bibnamefont
  {Hosono}}, \bibinfo {author} {\bibfnamefont {A.}~\bibnamefont {Yamamoto}},
  \bibinfo {author} {\bibfnamefont {H.}~\bibnamefont {Hiramatsu}}, \ and\
  \bibinfo {author} {\bibfnamefont {Y.}~\bibnamefont {Ma}},\ }\href {\doibase
  https://doi.org/10.1016/j.mattod.2017.09.006} {\bibfield  {journal} {\bibinfo
   {journal} {Materials Today}\ } (\bibinfo {year} {2017}),\
  https://doi.org/10.1016/j.mattod.2017.09.006}\BibitemShut {NoStop}%
\bibitem [{\citenamefont {Fu}\ \emph {et~al.}(2007)\citenamefont {Fu},
  \citenamefont {Kane},\ and\ \citenamefont {Mele}}]{fu2007titd}%
  \BibitemOpen
  \bibfield  {author} {\bibinfo {author} {\bibfnamefont {L.}~\bibnamefont
  {Fu}}, \bibinfo {author} {\bibfnamefont {C.~L.}\ \bibnamefont {Kane}}, \ and\
  \bibinfo {author} {\bibfnamefont {E.~J.}\ \bibnamefont {Mele}},\ }\href
  {\doibase 10.1103/PhysRevLett.98.106803} {\bibfield  {journal} {\bibinfo
  {journal} {Phys. Rev. Lett.}\ }\textbf {\bibinfo {volume} {98}},\ \bibinfo
  {pages} {106803} (\bibinfo {year} {2007})}\BibitemShut {NoStop}%
\bibitem [{\citenamefont {Zhang}\ \emph {et~al.}(2009)\citenamefont {Zhang},
  \citenamefont {Liu}, \citenamefont {Qi}, \citenamefont {Dai}, \citenamefont
  {Fang},\ and\ \citenamefont {Zhang}}]{hzhang2009}%
  \BibitemOpen
  \bibfield  {author} {\bibinfo {author} {\bibfnamefont {H.}~\bibnamefont
  {Zhang}}, \bibinfo {author} {\bibfnamefont {C.-X.}\ \bibnamefont {Liu}},
  \bibinfo {author} {\bibfnamefont {X.-L.}\ \bibnamefont {Qi}}, \bibinfo
  {author} {\bibfnamefont {X.}~\bibnamefont {Dai}}, \bibinfo {author}
  {\bibfnamefont {Z.}~\bibnamefont {Fang}}, \ and\ \bibinfo {author}
  {\bibfnamefont {S.-C.}\ \bibnamefont {Zhang}},\ }\href@noop {} {\bibfield
  {journal} {\bibinfo  {journal} {Nature Physics}\ }\textbf {\bibinfo {volume}
  {5}},\ \bibinfo {pages} {438} (\bibinfo {year} {2009})}\BibitemShut {NoStop}%
\bibitem [{\citenamefont {Zhang}\ \emph {et~al.}(2010)\citenamefont {Zhang},
  \citenamefont {Yu}, \citenamefont {Zhang}, \citenamefont {Dai},\ and\
  \citenamefont {Fang}}]{wzhang2010}%
  \BibitemOpen
  \bibfield  {author} {\bibinfo {author} {\bibfnamefont {W.}~\bibnamefont
  {Zhang}}, \bibinfo {author} {\bibfnamefont {R.}~\bibnamefont {Yu}}, \bibinfo
  {author} {\bibfnamefont {H.-J.}\ \bibnamefont {Zhang}}, \bibinfo {author}
  {\bibfnamefont {X.}~\bibnamefont {Dai}}, \ and\ \bibinfo {author}
  {\bibfnamefont {Z.}~\bibnamefont {Fang}},\ }\href@noop {} {\bibfield
  {journal} {\bibinfo  {journal} {New Journal of Physics}\ }\textbf {\bibinfo
  {volume} {12}},\ \bibinfo {pages} {065013} (\bibinfo {year}
  {2010})}\BibitemShut {NoStop}%
\bibitem [{\citenamefont {Kresse}\ and\ \citenamefont
  {Hafner}(1993)}]{kresse1993}%
  \BibitemOpen
  \bibfield  {author} {\bibinfo {author} {\bibfnamefont {G.}~\bibnamefont
  {Kresse}}\ and\ \bibinfo {author} {\bibfnamefont {J.}~\bibnamefont
  {Hafner}},\ }\href {\doibase 10.1103/PhysRevB.47.558} {\bibfield  {journal}
  {\bibinfo  {journal} {Phys. Rev. B}\ }\textbf {\bibinfo {volume} {47}},\
  \bibinfo {pages} {558} (\bibinfo {year} {1993})}\BibitemShut {NoStop}%
\bibitem [{\citenamefont {Kresse}\ and\ \citenamefont
  {Furthm\"uller}(1996)}]{kresse1996}%
  \BibitemOpen
  \bibfield  {author} {\bibinfo {author} {\bibfnamefont {G.}~\bibnamefont
  {Kresse}}\ and\ \bibinfo {author} {\bibfnamefont {J.}~\bibnamefont
  {Furthm\"uller}},\ }\href {\doibase 10.1103/PhysRevB.54.11169} {\bibfield
  {journal} {\bibinfo  {journal} {Phys. Rev. B}\ }\textbf {\bibinfo {volume}
  {54}},\ \bibinfo {pages} {11169} (\bibinfo {year} {1996})}\BibitemShut
  {NoStop}%
\bibitem [{\citenamefont {Bl\"ochl}(1994)}]{blochl1994}%
  \BibitemOpen
  \bibfield  {author} {\bibinfo {author} {\bibfnamefont {P.~E.}\ \bibnamefont
  {Bl\"ochl}},\ }\href {\doibase 10.1103/PhysRevB.50.17953} {\bibfield
  {journal} {\bibinfo  {journal} {Phys. Rev. B}\ }\textbf {\bibinfo {volume}
  {50}},\ \bibinfo {pages} {17953} (\bibinfo {year} {1994})}\BibitemShut
  {NoStop}%
\bibitem [{\citenamefont {Perdew}\ \emph {et~al.}(2008)\citenamefont {Perdew},
  \citenamefont {Ruzsinszky}, \citenamefont {Csonka}, \citenamefont {Vydrov},
  \citenamefont {Scuseria}, \citenamefont {Constantin}, \citenamefont {Zhou},\
  and\ \citenamefont {Burke}}]{perdew2008}%
  \BibitemOpen
  \bibfield  {author} {\bibinfo {author} {\bibfnamefont {J.~P.}\ \bibnamefont
  {Perdew}}, \bibinfo {author} {\bibfnamefont {A.}~\bibnamefont {Ruzsinszky}},
  \bibinfo {author} {\bibfnamefont {G.~I.}\ \bibnamefont {Csonka}}, \bibinfo
  {author} {\bibfnamefont {O.~A.}\ \bibnamefont {Vydrov}}, \bibinfo {author}
  {\bibfnamefont {G.~E.}\ \bibnamefont {Scuseria}}, \bibinfo {author}
  {\bibfnamefont {L.~A.}\ \bibnamefont {Constantin}}, \bibinfo {author}
  {\bibfnamefont {X.}~\bibnamefont {Zhou}}, \ and\ \bibinfo {author}
  {\bibfnamefont {K.}~\bibnamefont {Burke}},\ }\href {\doibase
  10.1103/PhysRevLett.100.136406} {\bibfield  {journal} {\bibinfo  {journal}
  {Phys. Rev. Lett.}\ }\textbf {\bibinfo {volume} {100}},\ \bibinfo {pages}
  {136406} (\bibinfo {year} {2008})}\BibitemShut {NoStop}%
\bibitem [{\citenamefont {Luo}\ \emph {et~al.}(1970)\citenamefont {Luo},
  \citenamefont {Vielhaber},\ and\ \citenamefont {Corenzwit}}]{luo1970}%
  \BibitemOpen
  \bibfield  {author} {\bibinfo {author} {\bibfnamefont {H.~L.}\ \bibnamefont
  {Luo}}, \bibinfo {author} {\bibfnamefont {E.}~\bibnamefont {Vielhaber}}, \
  and\ \bibinfo {author} {\bibfnamefont {E.}~\bibnamefont {Corenzwit}},\ }\href
  {\doibase 10.1007/BF01394489} {\bibfield  {journal} {\bibinfo  {journal}
  {Zeitschrift f{\"u}r Physik A Hadrons and nuclei}\ }\textbf {\bibinfo
  {volume} {230}},\ \bibinfo {pages} {443} (\bibinfo {year}
  {1970})}\BibitemShut {NoStop}%
\bibitem [{\citenamefont {Courtney}\ \emph {et~al.}(1965)\citenamefont
  {Courtney}, \citenamefont {Pearsall},\ and\ \citenamefont
  {Wulff}}]{courtney1965}%
  \BibitemOpen
  \bibfield  {author} {\bibinfo {author} {\bibfnamefont {T.~H.}\ \bibnamefont
  {Courtney}}, \bibinfo {author} {\bibfnamefont {G.~W.}\ \bibnamefont
  {Pearsall}}, \ and\ \bibinfo {author} {\bibfnamefont {J.}~\bibnamefont
  {Wulff}},\ }\href {\doibase 10.1063/1.1702960} {\bibfield  {journal}
  {\bibinfo  {journal} {Journal of Applied Physics}\ }\textbf {\bibinfo
  {volume} {36}},\ \bibinfo {pages} {3256} (\bibinfo {year}
  {1965})}\BibitemShut {NoStop}%
\bibitem [{\citenamefont {Furuseth}\ and\ \citenamefont
  {Kjekshus}(1964)}]{furuseth1964}%
  \BibitemOpen
  \bibfield  {author} {\bibinfo {author} {\bibfnamefont {S.}~\bibnamefont
  {Furuseth}}\ and\ \bibinfo {author} {\bibfnamefont {A.}~\bibnamefont
  {Kjekshus}},\ }\href@noop {} {\bibfield  {journal} {\bibinfo  {journal} {Acta
  Chem. Scand.}\ }\textbf {\bibinfo {volume} {18}},\ \bibinfo {pages} {1180}
  (\bibinfo {year} {1964})}\BibitemShut {NoStop}%
\bibitem [{\citenamefont {Guseva}\ \emph {et~al.}(1982)\citenamefont {Guseva},
  \citenamefont {Seropegin},\ and\ \citenamefont {Sokolovskaya}}]{guseva1982}%
  \BibitemOpen
  \bibfield  {author} {\bibinfo {author} {\bibfnamefont {I.}~\bibnamefont
  {Guseva}}, \bibinfo {author} {\bibfnamefont {Y.}~\bibnamefont {Seropegin}}, \
  and\ \bibinfo {author} {\bibfnamefont {E.}~\bibnamefont {Sokolovskaya}},\
  }\href {\doibase https://doi.org/10.1016/0022-5088(82)90047-9} {\bibfield
  {journal} {\bibinfo  {journal} {Journal of the Less Common Metals}\ }\textbf
  {\bibinfo {volume} {87}},\ \bibinfo {pages} {109 } (\bibinfo {year}
  {1982})}\BibitemShut {NoStop}%
\bibitem [{\citenamefont {Giannozzi}\ \emph {et~al.}(2009)\citenamefont
  {Giannozzi}, \citenamefont {Baroni}, \citenamefont {Bonini}, \citenamefont
  {Calandra}, \citenamefont {Car}, \citenamefont {Cavazzoni}, \citenamefont
  {Ceresoli}, \citenamefont {Chiarotti}, \citenamefont {Cococcioni},
  \citenamefont {Dabo}, \citenamefont {Corso}, \citenamefont {de~Gironcoli},
  \citenamefont {Fabris}, \citenamefont {Fratesi}, \citenamefont {Gebauer},
  \citenamefont {Gerstmann}, \citenamefont {Gougoussis}, \citenamefont
  {Kokalj}, \citenamefont {Lazzeri}, \citenamefont {Martin-Samos},
  \citenamefont {Marzari}, \citenamefont {Mauri}, \citenamefont {Mazzarello},
  \citenamefont {Paolini}, \citenamefont {Pasquarello}, \citenamefont
  {Paulatto}, \citenamefont {Sbraccia}, \citenamefont {Scandolo}, \citenamefont
  {Sclauzero}, \citenamefont {Seitsonen}, \citenamefont {Smogunov},
  \citenamefont {Umari},\ and\ \citenamefont {Wentzcovitch}}]{giannozzi2009}%
  \BibitemOpen
  \bibfield  {author} {\bibinfo {author} {\bibfnamefont {P.}~\bibnamefont
  {Giannozzi}}, \bibinfo {author} {\bibfnamefont {S.}~\bibnamefont {Baroni}},
  \bibinfo {author} {\bibfnamefont {N.}~\bibnamefont {Bonini}}, \bibinfo
  {author} {\bibfnamefont {M.}~\bibnamefont {Calandra}}, \bibinfo {author}
  {\bibfnamefont {R.}~\bibnamefont {Car}}, \bibinfo {author} {\bibfnamefont
  {C.}~\bibnamefont {Cavazzoni}}, \bibinfo {author} {\bibfnamefont
  {D.}~\bibnamefont {Ceresoli}}, \bibinfo {author} {\bibfnamefont {G.~L.}\
  \bibnamefont {Chiarotti}}, \bibinfo {author} {\bibfnamefont {M.}~\bibnamefont
  {Cococcioni}}, \bibinfo {author} {\bibfnamefont {I.}~\bibnamefont {Dabo}},
  \bibinfo {author} {\bibfnamefont {A.~D.}\ \bibnamefont {Corso}}, \bibinfo
  {author} {\bibfnamefont {S.}~\bibnamefont {de~Gironcoli}}, \bibinfo {author}
  {\bibfnamefont {S.}~\bibnamefont {Fabris}}, \bibinfo {author} {\bibfnamefont
  {G.}~\bibnamefont {Fratesi}}, \bibinfo {author} {\bibfnamefont
  {R.}~\bibnamefont {Gebauer}}, \bibinfo {author} {\bibfnamefont
  {U.}~\bibnamefont {Gerstmann}}, \bibinfo {author} {\bibfnamefont
  {C.}~\bibnamefont {Gougoussis}}, \bibinfo {author} {\bibfnamefont
  {A.}~\bibnamefont {Kokalj}}, \bibinfo {author} {\bibfnamefont
  {M.}~\bibnamefont {Lazzeri}}, \bibinfo {author} {\bibfnamefont
  {L.}~\bibnamefont {Martin-Samos}}, \bibinfo {author} {\bibfnamefont
  {N.}~\bibnamefont {Marzari}}, \bibinfo {author} {\bibfnamefont
  {F.}~\bibnamefont {Mauri}}, \bibinfo {author} {\bibfnamefont
  {R.}~\bibnamefont {Mazzarello}}, \bibinfo {author} {\bibfnamefont
  {S.}~\bibnamefont {Paolini}}, \bibinfo {author} {\bibfnamefont
  {A.}~\bibnamefont {Pasquarello}}, \bibinfo {author} {\bibfnamefont
  {L.}~\bibnamefont {Paulatto}}, \bibinfo {author} {\bibfnamefont
  {C.}~\bibnamefont {Sbraccia}}, \bibinfo {author} {\bibfnamefont
  {S.}~\bibnamefont {Scandolo}}, \bibinfo {author} {\bibfnamefont
  {G.}~\bibnamefont {Sclauzero}}, \bibinfo {author} {\bibfnamefont {A.~P.}\
  \bibnamefont {Seitsonen}}, \bibinfo {author} {\bibfnamefont {A.}~\bibnamefont
  {Smogunov}}, \bibinfo {author} {\bibfnamefont {P.}~\bibnamefont {Umari}}, \
  and\ \bibinfo {author} {\bibfnamefont {R.~M.}\ \bibnamefont {Wentzcovitch}},\
  }\href {http://stacks.iop.org/0953-8984/21/i=39/a=395502} {\bibfield
  {journal} {\bibinfo  {journal} {Journal of Physics: Condensed Matter}\
  }\textbf {\bibinfo {volume} {21}},\ \bibinfo {pages} {395502} (\bibinfo
  {year} {2009})}\BibitemShut {NoStop}%
\bibitem [{\citenamefont {Fu}\ and\ \citenamefont {Kane}(2007)}]{fu2007}%
  \BibitemOpen
  \bibfield  {author} {\bibinfo {author} {\bibfnamefont {L.}~\bibnamefont
  {Fu}}\ and\ \bibinfo {author} {\bibfnamefont {C.~L.}\ \bibnamefont {Kane}},\
  }\href {\doibase 10.1103/PhysRevB.76.045302} {\bibfield  {journal} {\bibinfo
  {journal} {Phys. Rev. B}\ }\textbf {\bibinfo {volume} {76}},\ \bibinfo
  {pages} {045302} (\bibinfo {year} {2007})}\BibitemShut {NoStop}%
\bibitem [{\citenamefont {Kim}\ \emph {et~al.}(2016)\citenamefont {Kim},
  \citenamefont {Ihm},\ and\ \citenamefont {Chung}}]{mskim2016}%
  \BibitemOpen
  \bibfield  {author} {\bibinfo {author} {\bibfnamefont {M.}~\bibnamefont
  {Kim}}, \bibinfo {author} {\bibfnamefont {J.}~\bibnamefont {Ihm}}, \ and\
  \bibinfo {author} {\bibfnamefont {S.~B.}\ \bibnamefont {Chung}},\ }\href
  {\doibase 10.1103/PhysRevB.94.115431} {\bibfield  {journal} {\bibinfo
  {journal} {Phys. Rev. B}\ }\textbf {\bibinfo {volume} {94}},\ \bibinfo
  {pages} {115431} (\bibinfo {year} {2016})}\BibitemShut {NoStop}%
\bibitem [{\citenamefont {Stroppa}\ \emph {et~al.}(2014)\citenamefont
  {Stroppa}, \citenamefont {Di~Sante}, \citenamefont {Barone}, \citenamefont
  {Bokdam}, \citenamefont {Kresse}, \citenamefont {Franchini}, \citenamefont
  {Whangbo},\ and\ \citenamefont {Picozzi}}]{stroppa2014}%
  \BibitemOpen
  \bibfield  {author} {\bibinfo {author} {\bibfnamefont {A.}~\bibnamefont
  {Stroppa}}, \bibinfo {author} {\bibfnamefont {D.}~\bibnamefont {Di~Sante}},
  \bibinfo {author} {\bibfnamefont {P.}~\bibnamefont {Barone}}, \bibinfo
  {author} {\bibfnamefont {M.}~\bibnamefont {Bokdam}}, \bibinfo {author}
  {\bibfnamefont {G.}~\bibnamefont {Kresse}}, \bibinfo {author} {\bibfnamefont
  {C.}~\bibnamefont {Franchini}}, \bibinfo {author} {\bibfnamefont {M.-H.}\
  \bibnamefont {Whangbo}}, \ and\ \bibinfo {author} {\bibfnamefont
  {S.}~\bibnamefont {Picozzi}},\ }\href@noop {} {\bibfield  {journal} {\bibinfo
   {journal} {Nat Comms}\ }\textbf {\bibinfo {volume} {5}},\ \bibinfo {pages}
  {5900} (\bibinfo {year} {2014})}\BibitemShut {NoStop}%
\bibitem [{\citenamefont {Ye}\ \emph {et~al.}(2010)\citenamefont {Ye},
  \citenamefont {Inoue}, \citenamefont {Kobayashi}, \citenamefont {Kasahara},
  \citenamefont {Yuan}, \citenamefont {Shimotani},\ and\ \citenamefont
  {Iwasa}}]{ye2010}%
  \BibitemOpen
  \bibfield  {author} {\bibinfo {author} {\bibfnamefont {J.~T.}\ \bibnamefont
  {Ye}}, \bibinfo {author} {\bibfnamefont {S.}~\bibnamefont {Inoue}}, \bibinfo
  {author} {\bibfnamefont {K.}~\bibnamefont {Kobayashi}}, \bibinfo {author}
  {\bibfnamefont {Y.}~\bibnamefont {Kasahara}}, \bibinfo {author}
  {\bibfnamefont {H.~T.}\ \bibnamefont {Yuan}}, \bibinfo {author}
  {\bibfnamefont {H.}~\bibnamefont {Shimotani}}, \ and\ \bibinfo {author}
  {\bibfnamefont {Y.}~\bibnamefont {Iwasa}},\ }\href@noop {} {\bibfield
  {journal} {\bibinfo  {journal} {Nature Materials}\ }\textbf {\bibinfo
  {volume} {9}},\ \bibinfo {pages} {125} (\bibinfo {year} {2010})}\BibitemShut
  {NoStop}%
\bibitem [{\citenamefont {Jeong}\ \emph {et~al.}(2013)\citenamefont {Jeong},
  \citenamefont {Aetukuri}, \citenamefont {Graf}, \citenamefont {Schladt},
  \citenamefont {Samant},\ and\ \citenamefont {Parkin}}]{jeong2013}%
  \BibitemOpen
  \bibfield  {author} {\bibinfo {author} {\bibfnamefont {J.}~\bibnamefont
  {Jeong}}, \bibinfo {author} {\bibfnamefont {N.}~\bibnamefont {Aetukuri}},
  \bibinfo {author} {\bibfnamefont {T.}~\bibnamefont {Graf}}, \bibinfo {author}
  {\bibfnamefont {T.~D.}\ \bibnamefont {Schladt}}, \bibinfo {author}
  {\bibfnamefont {M.~G.}\ \bibnamefont {Samant}}, \ and\ \bibinfo {author}
  {\bibfnamefont {S.~S.~P.}\ \bibnamefont {Parkin}},\ }\href {\doibase
  10.1126/science.1230512} {\bibfield  {journal} {\bibinfo  {journal}
  {Science}\ }\textbf {\bibinfo {volume} {339}},\ \bibinfo {pages} {1402}
  (\bibinfo {year} {2013})},\ \Eprint
  {http://arxiv.org/abs/http://science.sciencemag.org/content/339/6126/1402.full.pdf}
  {http://science.sciencemag.org/content/339/6126/1402.full.pdf} \BibitemShut
  {NoStop}%
\bibitem [{\citenamefont {Hsieh}\ \emph {et~al.}(2008)\citenamefont {Hsieh},
  \citenamefont {Qian}, \citenamefont {Wray}, \citenamefont {Xia},
  \citenamefont {Hor}, \citenamefont {Cava},\ and\ \citenamefont
  {Hasan}}]{hsieh2008}%
  \BibitemOpen
  \bibfield  {author} {\bibinfo {author} {\bibfnamefont {D.}~\bibnamefont
  {Hsieh}}, \bibinfo {author} {\bibfnamefont {D.}~\bibnamefont {Qian}},
  \bibinfo {author} {\bibfnamefont {L.}~\bibnamefont {Wray}}, \bibinfo {author}
  {\bibfnamefont {Y.}~\bibnamefont {Xia}}, \bibinfo {author} {\bibfnamefont
  {Y.~S.}\ \bibnamefont {Hor}}, \bibinfo {author} {\bibfnamefont {R.~J.}\
  \bibnamefont {Cava}}, \ and\ \bibinfo {author} {\bibfnamefont {M.~Z.}\
  \bibnamefont {Hasan}},\ }\href@noop {} {\bibfield  {journal} {\bibinfo
  {journal} {Nature}\ }\textbf {\bibinfo {volume} {452}},\ \bibinfo {pages}
  {970} (\bibinfo {year} {2008})}\BibitemShut {NoStop}%
\end{thebibliography}
\end{document}